\documentclass[sigconf,nonacm]{acmart}
\setcopyright{none}
\acmDOI{}
\acmISBN{}
\acmConference{}{}{}
\usepackage{placeins}
\usepackage{tabularx,array}

\newlength{\summarysep}
\newcounter{observation}
\newcommand{\observationtext}[2]{%
  \par\nopagebreak[4]\addvspace{\summarysep}\nopagebreak[4]
  \noindent\refstepcounter{observation}\begingroup
  \setlength{\fboxsep}{3.5pt}%
  \setlength{\fboxrule}{0.45pt}%
  \fbox{\begin{minipage}{\dimexpr\columnwidth-2\fboxsep-2\fboxrule\relax}
    \raggedright
    \noindent\textbf{Observation~\theobservation.} #2
  \end{minipage}}%
  \endgroup
  \par\addvspace{\summarysep}
}

\newcommand{\openquestiontext}[3][4]{%
  \par\nopagebreak[4]\addvspace{\summarysep}\nopagebreak[4]
  \noindent\begingroup
  \setlength{\fboxsep}{3.5pt}%
  \setlength{\fboxrule}{0.45pt}%
  \fbox{\begin{minipage}{\dimexpr\columnwidth-2\fboxsep-2\fboxrule\relax}
    \raggedright
    \noindent\textbf{Open question~#2.} #3
  \end{minipage}}%
  \endgroup
  \par\nopagebreak[#1]\addvspace{\summarysep}\nopagebreak[#1]
}
\makeatletter

\newcommand{\inlinefigurecaption}[2][3pt]{%
  \def\@captype{figure}%
  \small
  \caption{#2}%
  \normalsize
}

\newcommand{\inlinetablecaption}[1]{%
  \def\@captype{table}%
  \captionsetup{hypcap=false}%
  \small
  \caption{#1}%
  \normalsize
}
\makeatother

\newif\ifpreprintbibbalanced
\AtBeginEnvironment{thebibliography}{%
  \let\preprintoriginalbibitem\bibitem
  \renewcommand{\bibitem}{%
    \ifpreprintbibbalanced\else
      \ifnum\value{page}=\getpagerefnumber{TotPages}\relax
        \global\preprintbibbalancedtrue\balance
      \fi
    \fi
    \preprintoriginalbibitem
  }%
}

\hypersetup{hidelinks}

\begin{document}

\date{}
\title[Understanding Agent Serving at Production Scale]{Semantics, Workflows, and Infrastructure: Understanding Agent Serving at Production Scale}

\author{Yihao Zheng}
\affiliation[obeypunctuation=true]{\institution{The Chinese University of Hong Kong, Shenzhen}\city{}\country{}}
\author{Jingzhe Jiang}
\affiliation[obeypunctuation=true]{\institution{The Chinese University of Hong Kong, Shenzhen}\city{}\country{}}

\author{Dejiang Zhu}
\affiliation[obeypunctuation=true]{\institution{Ant Group}\city{}\country{}}

\author{Zhiyuan Tan}
\affiliation[obeypunctuation=true]{\institution{The Chinese University of Hong Kong, Shenzhen}\city{}\country{}}

\author{Yang Tian}
\affiliation[obeypunctuation=true]{\institution{Ant Group}\city{}\country{}}

\author{Tao Wang}
\authornote{Corresponding authors.}
\affiliation[obeypunctuation=true]{\institution{Ant Group}\city{}\country{}}

\author{Minchen Yu}
\authornotemark[1]
\affiliation[obeypunctuation=true]{\institution{The Chinese University of Hong Kong, Shenzhen}\city{}\country{}}

\renewcommand{\shortauthors}{Zheng et al.}

\begin{abstract}

Large language model (LLM) agents execute applications through a workflow of inference requests with tool calls and user interactions. Serving these applications at production scale requires understanding how application behavior shapes inference demand and for guiding efficient execution.
Recent characterization studies provide request-level workload measurements and agent execution analysis. However, an end-to-end view connecting task initiation, workflow execution, and inference infrastructure remains unexplored.
In this paper, we analyze a two-week trace of 11.7 million requests from a large-scale production platform for general-purpose agents, backed by inference infrastructure comprising over 10k GPUs.
We characterize the platform at three connected levels: task-level initiation semantics, workflow-level execution patterns, and infrastructure-level serving demands.
Our measurements reveal workload patterns such as highly skewed request volumes across sessions, rare execution overlap among logical sibling requests, and context reuse across task boundaries.
Building on these observations, we analyze deployment implications and identify open problems to guide future research on agent serving systems.

\end{abstract}

\maketitle

\label{sourcepara:1}
\section{Introduction}

\label{sourcepara:2}
Large language model (LLM) agents are increasingly used to automate
real-world activities, including programming, data analysis, and office
work~\cite{yang2024sweagent,drouin2024workarena}.
As illustrated in Fig.~\label{first:workflow_hierarchy}\ref{fig:workflow_hierarchy}, an agent framework
maintains long-lived \emph{sessions}, each preserving a shared context
across one or more \emph{tasks}.
A task is a unit of work defined by a user-specified goal, such as
generating code or testing an implementation.
Tasks within the same session can draw on context accumulated from
earlier interactions.
Executing a task may generate multiple LLM inference requests interleaved with tool calls.
The workflow structure captures the relationships among these requests within and across tasks.

\begin{figure}[t]
\centering
\includegraphics[width=\columnwidth]{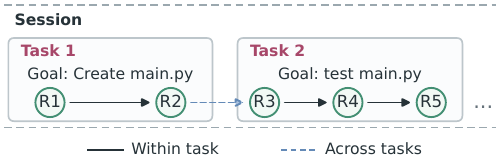}
\setcounter{figure}{0}
\caption{Three-level schematic of sessions, tasks, and requests. A session is a long-running conversation with historical context. Each task pursues a specific goal and may involve multiple LLM requests and tool calls.}
\label{fig:workflow_hierarchy}
\Description{Schematic of a session containing tasks, with inference requests and tool calls connected within and across tasks.}
\end{figure}

\label{sourcepara:3}
Agentic applications are becoming a major workload for LLM inference infrastructure~\cite{anthropic2026cadences,openrouter2026agentic}.
Production workload characterization and public datasets provide an empirical basis for designing and evaluating serving systems~\cite{xiang2026servegen,wang2025burstgpt}.
Recent studies characterize execution patterns, token consumption, and cache behavior in coding agents such as GitHub Copilot~\cite{liu2026copilot}, Claude Code, and Codex~\cite{zhu2026tracelab}.
However, how task initiation and workflow execution shape inference demand remains insufficiently characterized.
Addressing this gap requires an \emph{end-to-end} view connecting task- and workflow-level semantics with serving infrastructure.
Such a view can clarify how request load, execution patterns, and context reuse vary with task and workflow behavior, informing serving optimizations such as request scheduling and state management.

\label{sourcepara:4}
In this paper, we present a comprehensive characterization of agentic workloads in a large-scale production platform.
Our study uses a two-week trace containing 11.7M model requests and 948.4k sessions.
The workloads span coding, testing, data analysis, office work, and automated operations.
The platform deploys multiple open-source LLMs as inference backends for agent services on a cluster of over 10k GPUs.
The collected trace data provide information across multiple system layers, covering agent execution, inference requests, and resource utilization of serving infrastructure.

\label{sourcepara:5}
We organize our characterization at three connected levels:
the task level, the workflow level, and the infrastructure level.
At the \emph{task level}, we compare user-triggered and agent-triggered tasks
and analyze recurring activity, examining how task initiation patterns
relate to workload volume and its variation over time.
At the \emph{workflow level}, we reconstruct workflow relationships within
and across tasks as directed acyclic graphs (DAGs), and examine how their structure relates to request load and execution patterns.
At the \emph{infrastructure level}, we quantify how resource demands and context residency are distributed across sessions.
We then examine context reuse within and across tasks to study their implications for request
placement and state management.
Table~\ref{tab:agent-trace-comparison} compares the scope of our analysis
with prior agent-workload studies.
We highlight the following observations at each level.

\label{sourcepara:6}
\noindent\textbf{Task-level initiation semantics.}
We characterize user-triggered and
agent-triggered tasks by comparing request counts, context
lengths, and activity spans.
To connect individual task initiations to aggregate workload, we
analyze request concentration across sessions with different trigger
compositions and examine daily activity and recurrence intervals
within groups of repeated automated work.
Our analysis yields four observations:
(i) Agent-triggered tasks account for 35.8\% of tasks but carry
lighter per-task workloads than user-triggered tasks.
(ii) Request volume is highly concentrated: the top 5\% of sessions
by request count account for 56.1\% of requests, with sessions
containing only user-triggered tasks contributing most requests
in this group.
(iii) User-triggered workloads combine workday rhythms with greater
variation in task size and arrival timing, whereas agent-triggered
activity is more evenly distributed throughout the day.
(iv) Recurring automated activity concentrates in frequently
activated task groups: groups with median recurrence intervals
of at most 30 minutes contain 88.3\% of recurring tasks.

\label{sourcepara:7}
\noindent\textbf{Workflow-level execution patterns.}
For agent workflows, we measure graph size, depth, and width, and combine structural
analysis with request service intervals and inter-request gaps
to distinguish workflow organization from observed execution.
We further examine waiting and successor cache reuse along
tool-associated request transitions.
This analysis yields five observations:
(i) Most workflow DAGs are small and narrow, yet a minority of
sessions with request-graph width greater than one accounts for
most requests.
(ii) Tasks with serial request structures account for 77.6\% of
requests, despite comprising less than half of all tasks.
(iii) In 12.0\% of serial tasks, one inter-request gap exceeds half the task span; these tasks have longer spans on average.
(iv) Logical branching rarely corresponds to overlapping request
execution: only 0.225\% of logical sibling request pairs have
overlapping service intervals.
(v) Waiting-time and successor cache-reuse distributions differ
across tool categories.

\label{sourcepara:8}
\noindent\textbf{Infrastructure-level serving demands.}
We examine serving demand beyond request counts by measuring
new-prefill tokens, decode tokens, and logical context residency,
which captures both context size and retention duration.
We relate observed cache reuse to task transitions, context changes,
and idle intervals to examine where serving affinity persists or
changes.
These analyses yield three observations:
(i) The same small fraction of sessions accounts for most token
work and logical context residency; within this group, the decode
phase accounts for most cumulative request service time.
(ii) Task boundaries need not align with serving affinity:
context reuse can span tasks, while context changes can occur
within a task.
We therefore introduce \emph{episodes} as serving-affinity units
that group requests continuing a historical context within a
retention horizon.
(iii) An offline retention model shows that longer waiting-time
tails need not warrant longer cache lifetimes, and that retention
priorities across waiting types can change with the retention budget.

\label{sourcepara:9}
\begin{table}[!t]
\centering
\begin{minipage}{\columnwidth}
  \centering
  \inlinetablecaption{Comparison of agent-workload Characterizations. }
  \label{tab:agent-trace-comparison}
  \begingroup
  \fontsize{8.5}{10.0}\selectfont
  \setlength{\tabcolsep}{2.5pt}
  \renewcommand{\arraystretch}{1.14}
  \renewcommand{\tabularxcolumn}[1]{m{#1}}
  \setlength{\aboverulesep}{1.5pt}
  \setlength{\belowrulesep}{1.5pt}
  \begin{tabularx}{\linewidth}{@{}m{0.30\linewidth}*{3}{>{\centering\arraybackslash}X}@{}}
  \toprule[0.7pt]
  & \textbf{Copilot~\cite{liu2026copilot}} & \textbf{TraceLab~\cite{zhu2026tracelab}} & \textbf{Ours} \\
  \midrule[0.35pt]
  Workload & Coding & Coding & General \\
  Duration & 1 week & \(\sim\)8 months & 2 weeks \\
  Requests & 760.5M & 357.2k & 11.69M \\
  Sessions & 13.5M & 4.3k & 948.4k \\
  GPUs & \textendash & \textendash & \(O(10\mathrm{k})\) \\
  \midrule[0.35pt]
  \multicolumn{4}{@{}l@{}}{\bfseries Task-level initiation semantics (\S\ref{sec:task-initiation})} \\
  \midrule[0.3pt]
  Triggers & \(\checkmark\) & \(\times\) & \(\checkmark\) \\
  Recurrence & \(\times\) & \(\times\) & \(\checkmark\) \\
  \midrule[0.35pt]
  \multicolumn{4}{@{}l@{}}{\bfseries Workflow-level execution patterns (\S\ref{sec:workflow-structure-execution})} \\
  \midrule[0.3pt]
  Transitions \& tools & \(\checkmark\) & \(\checkmark\) & \(\checkmark\) \\
  Workflow DAGs & \(\times\) & \(\times\) & \(\checkmark\) \\
  \midrule[0.35pt]
  \multicolumn{4}{@{}l@{}}{\bfseries Infrastructure-level serving demands (\S\ref{sec:runtime-performance})} \\
  \midrule[0.3pt]
  Reuse \& retention & \(\checkmark\) & \(\checkmark\) & \(\checkmark\) \\
  Serving granularity & \(\times\) & \(\times\) & \(\checkmark\) \\
  \bottomrule[0.7pt]
  \end{tabularx}
  \endgroup
\end{minipage}
\end{table}

\label{sourcepara:10}
\noindent\textbf{Implications and open questions.}
Building on these findings, we discuss three directions for agent serving.
(i) \emph{Cross-layer abstraction.}
Task initiation patterns and workflow dependencies could guide capacity
allocation and request scheduling.
This raises the question of what abstractions are needed to exchange
task, workflow, and resource information between agent runtimes and
serving engines.
(ii) \emph{Episode-based scheduling.}
Episodes offer an alternative to binding an entire session to one
serving instance.
An open question is how remaining-work estimates can guide request
scheduling and reassignment, weighing load-balancing benefits against
the cost of transferring or recomputing cached context.
(iii) \emph{Adaptive KV cache management.}
Our offline model shows that adaptively configuring cache lifetimes across workloads
can reduce recomputation compared with a uniform lifetime setup.
This motivates online policies that adjust cache lifetimes based on
expected return times, context reuse, and available memory.
By sharing these findings and open questions from our production
experience, we hope to facilitate future research on agent serving systems.

\begin{figure*}[t]
\centering
\includegraphics[width=\textwidth]{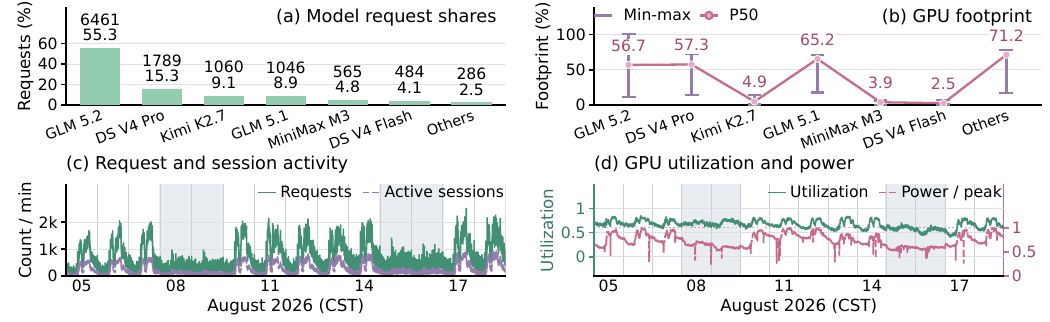}
\setcounter{figure}{1}
\caption{Platform workload overview. (a) Request shares and counts (k); (b) GPU footprint ranges and medians relative to the largest individual model peak. Others combines three models. (c) Requests and active sessions per minute. (d) GPU utilization and power relative to its two-week peak, on vertically separated scales. Shading marks weekends.}
\label{fig:platform_overview}
\Description{Four panels show request share by model, normalized GPU footprint by model, requests and active sessions over time, and normalized GPU utilization and power over time.}
\end{figure*}

\label{sourcepara:11}
\section{Background}

\label{sourcepara:12}
\subsection{Agent Workflow Execution}
\label{sec:agent-session-abstraction}

\label{sourcepara:13}
Agent frameworks organize LLM inference requests and tool calls into
workflows~\cite{wu2024autogen,zhuge2024gptswarm,zhang2025aflow}.
We describe agent execution using three units: \emph{sessions},
\emph{tasks}, and \emph{LLM inference requests}, as illustrated in
Fig.~\ref{fig:workflow_hierarchy}.
A \emph{session} maintains shared context across one or more tasks
and can persist between task executions.
This context includes interaction history, such as user messages,
model outputs, and tool results, that subsequent tasks can draw on.
A \emph{task} is a unit of work defined by a specific goal, such as
generating code or testing an implementation.
Tasks can begin through user interaction or automated activation;
either category can proceed through autonomous model--tool interactions.
An \emph{LLM inference request} is a single model invocation that
processes an input context and generates output.
Executing a task may involve multiple requests, interleaved with
tool calls and further user interaction.

\label{sourcepara:14}
Workflow structure captures the relationships among requests within
and across tasks.
Requests may extend an earlier context through serial execution or
branch into separate continuations.
These relationships evolve as model outputs, tool results, and user
feedback guide subsequent actions.
Workflow structure describes how work is organized, but does not
alone determine when requests execute: tool execution and external
waiting can separate successive requests.

\label{sourcepara:15}
\subsection{Workload Characterization}
\label{sec:workload-studies-background}

\label{sourcepara:16}
Production workload characterization provides an empirical basis for
understanding inference demand and designing and evaluating serving
systems.
BurstGPT characterizes request arrivals and demand
distributions~\cite{wang2025burstgpt}, while ServeGen examines
client heterogeneity to generate representative serving
workloads~\cite{xiang2026servegen}.
These studies support workload modeling through measurements of
request duration and input/output token lengths.
Agent workload studies further discuss agent
execution patterns. 
The Copilot study characterizes workflow patterns, resource
concentration, and cache behavior~\cite{liu2026copilot}.
TraceLab examines autonomous execution, context growth, and tool
calls in Claude Code and Codex~\cite{zhu2026tracelab}.
Both studies analyze transitions between requests and its implications for Context management.
CacheWise uses measurements of coding-agent workloads to guide
KVCache management~\cite{tiwari2026cachewise}.

\label{sourcepara:17}
Our study complements these efforts with an end-to-end view connecting
task-level initiation semantics, workflow-level execution patterns,
and infrastructure-level serving demand.
We examine how task initiation and recurring relate to
workloads, how request load and execution time vary across
workflow structures, and how context continuity and waiting inform
request placement and state management.
Table~\ref{tab:agent-trace-comparison} compares the scope of these
studies with ours.

\begin{figure}[t]
\centering
\includegraphics[width=\columnwidth]{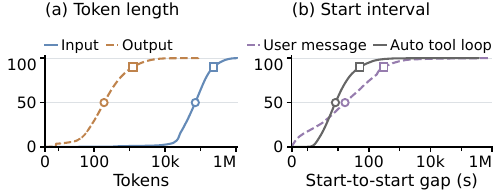}
\setcounter{figure}{2}
\caption{Request profiles. (a) Input/output lengths (tokens). Input P50/P90: 70k / 229k; Output P50/P90: 188 / 1.23k. (b) Start intervals (s). User message P50/P90: 16.95 / 309; Auto tool loop P50/P90: 8.29 / 51.45. Both panels show ECDFs (\%).}
\label{fig:request_profiles}
\Description{Two empirical cumulative distributions compare input and output token lengths, and request start intervals for user messages versus automated tool loops.}
\end{figure}

\begin{figure*}[t]
\centering
\includegraphics[width=\textwidth]{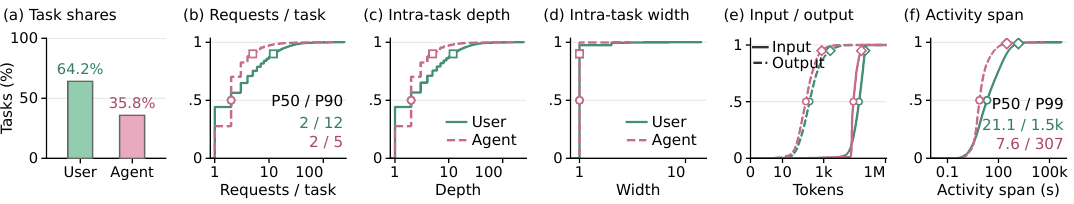}
\setcounter{figure}{3}
\caption{Task prevalence, request profiles, and activity spans differ by trigger. (a) Trigger shares; (b) Requests per task; (c,d) Depth and width of intra-task request DAGs by trigger; (e) Input/output lengths; (f) Task activity spans. Solid/dashed lines in (e) show input/output. ECDFs weight tasks (b--d,f) and requests (e) equally. Circles/squares mark P50/P90 in (b--d); circles/diamonds mark P50/P95 in (e) and P50/P99 in (f).}
\label{fig:trigger_profiles}
\Description{Six panels compare user-triggered and agent-triggered tasks in task share, requests per task, request-graph depth and width, token lengths, and activity span.}
\end{figure*}

\label{sourcepara:18}
\section{Workload Overview}
\label{sec:workload-overview}

\label{sourcepara:19}
\noindent\textbf{Platform and model demand.}
\label{sec:overview-setting}
As a large-scale agent service provider, our production platform serves frontier open-source LLMs such as
GLM 5.2 and DeepSeek V4 Pro and comprises over 1,000 model instances and over 10,000 GPUs. 
We analyze a two-week sample of production workloads containing
11.69 million requests, 1.42 million tasks, and 948.4k sessions.
The reported statistics in Table~\ref{tab:agent-trace-comparison} refer to this sampled trace~\footnote{For confidentiality reasons, we do not disclose
exact cluster sizes and total workload volumes.}. 
GPU metrics come from cluster-wide measurements: footprint and power
are normalized for presentation.
Request demand is concentrated in four model groups, which account for
88.58\% of requests (Figure~\label{first:platform_overview}\ref{fig:platform_overview}(a)),
but request shares alone do not describe GPU footprints directly.
Kimi K2.7 Code and GLM 5.1 receive similar request shares (9.06\% and 8.94\%), yet
their median GPU footprints differ by over 13-fold: 4.9\% versus 65.2\%
(Figure~\ref{fig:platform_overview}(b)). Footprint ranges and medians are normalized to the largest
individual model peak.

\label{sourcepara:20}
\noindent\textbf{Activity over time.}
\label{sec:overview-activity}
Workload activity exhibits daily and weekly rhythms, with larger
fluctuations than GPU power and utilization.
Request arrivals and session activity are higher during weekday daytime
and lower at night and on weekends
(Figure~\ref{fig:platform_overview}(c)). Session activity counts
distinct sessions with at least one request or task start in a minute.
During weekday nights, request arrivals fall to 17.6\% of their daytime
level, while GPU power remains at 80.0\% and utilization decreases from
74.8\% to 61.4\% (Figure~\ref{fig:platform_overview}(d)).
Section~\ref{sec:task-initiation} examines how temporal activity differs
by task triggering semantics.

\label{sourcepara:21}
\noindent\textbf{Request characteristics.}
Request profiles show long input contexts, short outputs, and substantial
cached input.
Median input and output lengths are 70,026 and 188 tokens, respectively
(Figure~\label{first:request_profiles}\ref{fig:request_profiles}(a)); both distributions have
long tails.
Request progression differs between user messages and auto tool loops:
the median start-to-start intervals are 16.95 and 8.29 seconds,
respectively (Figure~\ref{fig:request_profiles}(b)).
The median cached-input fraction is 98.1\%, while 6.34\% of requests have
no cached input.
Sections~\ref{sec:workflow-structure-execution}
and~\ref{sec:runtime-performance} examine request execution and reuse
through workflow dependencies and context continuity.

\begin{figure}[t]
\centering
\includegraphics[width=\columnwidth]{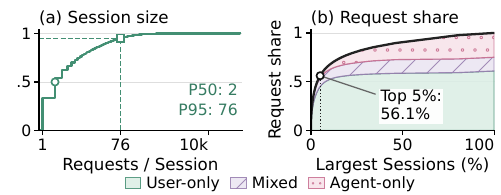}
\setcounter{figure}{4}
\caption{Session size and request concentration. (a) Requests per session, with P50/P95. (b) Cumulative share of all requests over sessions ranked from largest to smallest, split by trigger composition. Panel (a) shows a session-weighted ECDF.}
\label{fig:session_volume}
\Description{Session request volume is shown as a cumulative distribution and by ranked-session groups with different trigger compositions.}
\end{figure}

\label{sourcepara:22}
\section{Task-level Initiation Semantics}
\label{sec:task-initiation}
\label{sec:agentic-workload-analysis}
Task-level initiation semantics describe how work begins. User-triggered
tasks originate from user interaction, whereas agent-triggered tasks
originate from automated activation~\cite{microsoft2026eventtriggers}. This distinction concerns task
initiation, not autonomous continuation within a task. We examine the
prevalence and workload profiles of these categories, how their requests
accumulate across sessions, and task initiation timing.

\label{sourcepara:23}
\noindent\textbf{Trigger prevalence and workload profiles.}
User-triggered tasks are more prevalent, but automated activation also
contributes substantially: the two categories account for 64.17\% and
35.83\% of tasks (Figure~\label{first:trigger_profiles}\ref{fig:trigger_profiles}(a)). Both have
a P50 of two requests per task, yet user-triggered tasks include more
single-request cases and a longer upper tail: P90/P99 reach 12/39
requests, compared with 5/16 for agent-triggered tasks
(Figure~\ref{fig:trigger_profiles}(b)). The shared median therefore
hides a wider range of request counts in user-triggered work.
User-triggered tasks also have a longer upper tail in request
dependency-chain depth (Figure~\ref{fig:trigger_profiles}(c)).

\label{sourcepara:24}
Requests in user-triggered tasks carry more context and generate more
output. Median input lengths are 52.0k versus 27.4k tokens, and median
output lengths are 206 versus 136 tokens
(Figure~\ref{fig:trigger_profiles}(e)). Task activity persists longer
as well: medians are 21.1 versus 7.6 seconds, and P99 reaches 24.5 versus
5.1 minutes (Figure~\ref{fig:trigger_profiles}(f)).

\label{sourcepara:25}
Task counts miss differences in the context length,
and activity duration associated with each initiation.
\par\nopagebreak[2]
\observationtext{1}{Agent-triggered tasks account for 35.8\% of tasks but carry lighter per-task workloads than user-triggered tasks.}
\par

\begin{figure}[t]
\centering
\includegraphics[width=\columnwidth]{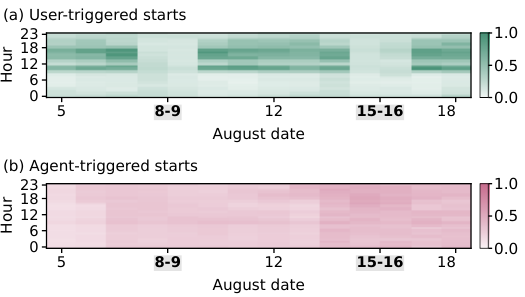}
\setcounter{figure}{5}
\caption{Within-trigger task-start shares (\%) with boxed weekends: (a) user-triggered tasks; (b) agent-triggered tasks.}
\label{fig:trigger_rhythm}
\Description{Heatmaps compare the daily activity patterns of user-triggered and agent-triggered work.}
\end{figure}

\label{sourcepara:26}
\noindent\textbf{Work volume and concentration.}
Sessions accumulate requests across tasks. User-only sessions contain
only user-triggered tasks, agent-only sessions contain only
agent-triggered tasks, and mixed sessions contain both.
Typical sessions are small, but their size distribution has a long tail:
requests per session have a median of two, with P95/P99 reaching 76/250
(Figure~\label{first:session_volume}\ref{fig:session_volume}(a)). Request volume is
also concentrated: the top 5\% and top 10\% of sessions contribute
56.1\% and 67.9\% of all requests, respectively
(Figure~\ref{fig:session_volume}(b)). Small sessions are
common, but they do not represent where most requests are generated.

\label{sourcepara:27}
User-only sessions dominate this high-volume tail. Their request share
rises from 61.0\% overall to 77.5\% within the top 10\% and 80.1\%
within the top 5\%. User-only sessions therefore contribute more
strongly to the high-volume tail than their overall request share
would suggest.

\label{sourcepara:28}
\par\nopagebreak[2]
\observationtext{2}{The top 5\% of sessions by request count account for 56.1\% of requests, with user-only sessions contributing most requests in this group.}
\par

\label{sourcepara:29}
\noindent\textbf{Activity rhythms and workload variability.}
Daily task-start profiles describe aggregate activity over time.
User-triggered activity exhibits recurring workday rhythms, whereas
agent-triggered activity is more evenly distributed over the day
(Figure~\label{first:trigger_rhythm}\ref{fig:trigger_rhythm}(a,b)).

\begin{figure}[t]
\centering
\includegraphics[width=\columnwidth]{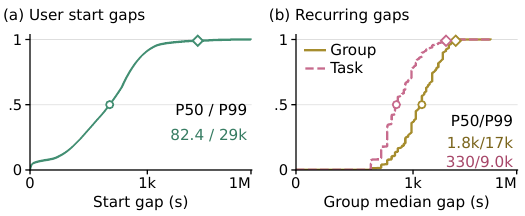}
\setcounter{figure}{6}
\caption{(a) User start gaps; (b) Recurring-group median gaps, group or task weighted. Circles/diamonds mark P50/P99. Both panels show ECDFs.}
\label{fig:trigger_gaps}
\Description{Two panels summarize recurrence intervals and the concentration of recurring automated tasks.}
\end{figure}

\label{sourcepara:30}
For user-triggered work, consecutive task starts within a session describe
the spacing of user initiations. These intervals combine closely spaced
starts with a long tail: the P50 gap is 82.4 seconds and the P99 is
8.07 hours (Figure~\label{first:trigger_gaps}\ref{fig:trigger_gaps}(a)). Workday rhythms thus
coexist with substantial variation in the intervals between
user-triggered tasks.

\label{sourcepara:31}
User-triggered tasks also vary in how many requests they generate and
how long their activity persists (Figure~\ref{fig:trigger_profiles}(b,f)).
Their request input lengths cover a wider range: P25/P75 are
32.2k/76.5k tokens, compared with 23.3k/38.0k for agent-triggered tasks
(Figure~\ref{fig:trigger_profiles}(e)). These differences affect
both the number of inference requests and the context each request
carries, adding variation beyond daily arrivals.
\par\nopagebreak[2]
\observationtext{3}{User-triggered workloads follow workday rhythms and show greater variation in task size and arrival temporal pattern than agent-triggered workloads, whose activity is more evenly distributed throughout the day.}
\par

\label{sourcepara:32}
For agent-triggered work, \emph{recurring task groups} describe repeated
instances of the same automated work in a Group. Groups with median gaps of at most
30 minutes contain 88.3\% of recurring tasks. Weighting groups by their
task counts shifts the median of gaps from 30 to 5.5 minutes
compared with equal group weighting
(Figure~\ref{fig:trigger_gaps}(b)). Short-interval groups therefore
contribute more heavily to recurring activity.
\par\nopagebreak[2]
\observationtext{4}{Recurring automated activity concentrates in task groups with median recurrence intervals of at most 30 minutes, which contain 88.3\% of recurring tasks.}
\par

\label{sourcepara:33}
\section{Workflow-level Execution Patterns}
\label{sec:workflow-structure-execution}

\label{sourcepara:34}
Workflow structure connects task initiation to the orchestration and timing
of requests. We examine graph scale and composition, relate structure
to trigger categories, then study load, execution timing, and request
transitions.

\label{sourcepara:35}
\subsection{Structural Organization and Volume}
\label{sec:structural-organization-load}

\begin{figure}[t]
\centering
\includegraphics[width=\columnwidth]{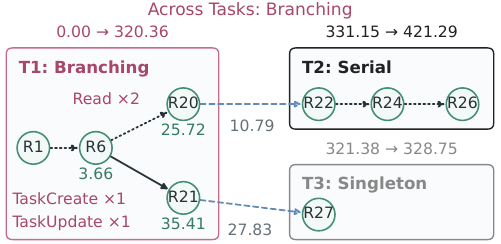}
\setcounter{figure}{7}
\caption{Workflow DAG example showing inter- and intra-task structures. Times are in seconds; green/gray values show request durations/end-to-start gaps. Pink labels show selected added-context tools (Read $\times2$: \mbox{R6--R7}).}
\label{fig:workflow_example}
\Description{A workflow example shows how request-level dependencies form task-level and session-level directed acyclic graphs.}
\end{figure}

\begin{figure*}[t]
\centering
\includegraphics[width=\textwidth]{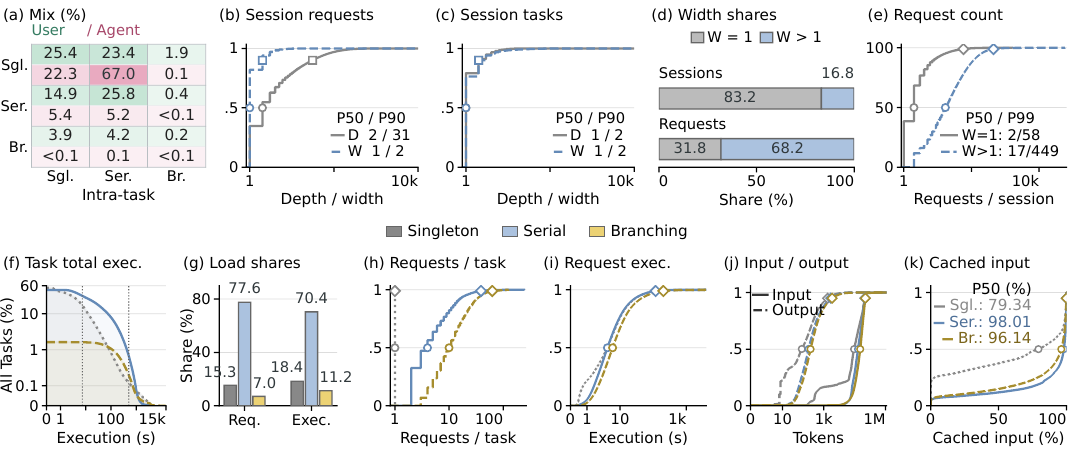}
\setcounter{figure}{8}
\caption{Workflow structure and request profiles. (a) Task shares by inter-task role (rows) and intra-task structure (columns), normalized within each trigger (user upper, agent lower). (b) Depth/width of request DAGs within sessions; (c) Depth/width of inter-task DAGs within sessions. (d) Session and request shares for $W=1$ and $W>1$; (e) request-count ECDFs for the same width groups, shown by solid and dashed lines, respectively. Workload by intra-task structure: (f) Share of all tasks with a given structure and cumulative execution above the threshold; (g) Shares of requests and cumulative execution duration; (h) Requests per task; (i) Request execution duration by intra-task structure; (j,k) Request input/output lengths and cached-input fraction by intra-task structure. ECDFs weight sessions (b,c,e), tasks (h), and requests (i--k) equally. Markers show P50/P90 (b,c), P50/P99 (e,h,i), and P50/P95 (j,k); dotted lines in (f) show P50/P99.}
\label{fig:structure_summary}
\Description{Eleven panels connect workflow structure, trigger composition and session graph widths with the request workload of singleton, serial and branching tasks.}
\end{figure*}

\label{sourcepara:36}
\noindent\textbf{Workflow structure and composition.}
\label{sec:trigger-structure-composition}
\label{sec:trigger-task-geometry}
We examine workflow orchestration at three levels: request DAGs within
sessions, task DAGs within sessions, and request DAGs within tasks. The
three levels workflow pattern describe how requests and tasks are organized
across a session;

\label{sourcepara:37}
Request edges represent transitions between requests and their logical
relationship within session- and task-level workflow DAGs. Task edges
represent context continuation, workflow invocations, or sub-agent
invocations across task boundaries. These graphs describe workflow
organization; request timestamps describe timing of execution.

\label{sourcepara:38}
Each task has an intra-task structure and an inter-task role. Singleton,
serial, and branching describe these two aspects.
Within a task, singleton means one request, branching contains a request
with multiple successors, and serial typically forms a request chain.
Among tasks, singleton denotes an isolated task; branching includes a
task with multiple successors and its direct successors; serial denotes
tasks in a task chain.

\label{sourcepara:39}
Figure~\label{first:workflow_example}\ref{fig:workflow_example} illustrates how
these structures combine. Within T1, the request graph forks at R6;
T2 contains a serial request chain, and T3 contains one request.
Across tasks, T1 connects to both T2 and T3. All three therefore have
branching inter-task roles, while their intra-task structures are
branching, serial, and singleton, respectively. A task's inter-task role
does not specify its internal request structure.

\label{sourcepara:40}
Size, depth, and width describe request and task graphs within sessions
and request graphs within tasks. For a graph $G=(V,E)$,
size $N=|V|$ counts nodes. Roots have layer $\ell(v)=0$; other node
layers follow the recurrence below. Depth $D$ counts nodes on the longest
dependency chain; width $W$ counts nodes in the largest layer, including
disconnected components.
\begingroup
\setlength{\abovedisplayskip}{0pt}
\setlength{\belowdisplayskip}{0pt}
\setlength{\abovedisplayshortskip}{0pt}
\setlength{\belowdisplayshortskip}{0pt}
\[
\begin{array}{c}
\ell(v)=1+\max_{u:(u,v)\in E}\ell(u),\\[3pt]
D=1+\max_v\ell(v),\quad W=\max_k|\{v:\ell(v)=k\}|.
\end{array}
\]
\endgroup

\begin{figure*}[t]
\centering
\includegraphics[width=\textwidth]{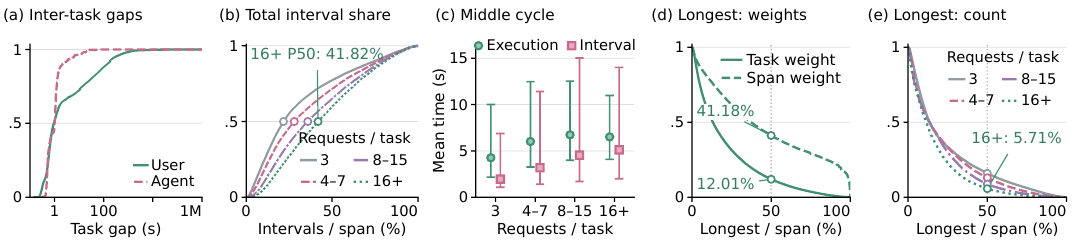}
\setcounter{figure}{9}
\caption{Workflow execution. (a) Inter-task gaps along serial chains, by successor trigger. Within serial tasks: (b) Cumulative request interval duration / task span; (c) Per-task middle-cycle execution/interval means (medians/IQRs); (d,e) Longest interval / task span, weighted by task or span in (d) and grouped by requests/task in (e). Panels (a,b) show ECDFs; (d,e) show CCDFs.}
\label{fig:serial_timing}
\Description{Five panels show continuation gaps between tasks, interval share within tasks, middle-cycle timing, and longest-interval shares weighted by tasks, spans, and request counts.}
\end{figure*}

\label{sourcepara:41}
Singleton and serial dominate the structural combinations. Tasks with
singleton or serial structures at both levels account for 89.46\% of user-triggered
tasks and 99.82\% of agent-triggered tasks
(Figure~\label{first:structure_summary}\ref{fig:structure_summary}(a)). Within sessions, request and task DAGs have median depths of 2 and
1, respectively, and both have median width one
(Figure~\ref{fig:structure_summary}(b,c)). Request graphs have
unit width in 82.10\% of sessions (Figure~\ref{fig:structure_summary}(b)); unit width also dominates
within-task request graphs in both trigger categories (Figure~\ref{fig:trigger_profiles}(d)).
Typical workflows have few dependency layers and narrow
widths. Larger graphs remain in the tail: session request and task DAGs
have P99 depths of 208 and eight, and P99 widths of five and 17,
respectively (Figure~\ref{fig:structure_summary}(b,c)). Within tasks, P99 request depth reaches
38 for user-triggered tasks and 16 for agent-triggered tasks (Figure~\ref{fig:trigger_profiles}(c)).

\label{sourcepara:42}
\noindent\textbf{Request concentration across sessions.}
Request volume concentrates in sessions with wider request graphs.
Sessions with width $W>1$ account for 16.78\% of sessions but carry
68.19\% of requests
(Figure~\ref{fig:structure_summary}(d)). Their median
request count is 17, compared with 2 for $W=1$ (Figure~\ref{fig:structure_summary}(e)), so the
size difference extends to typical sessions in each group. Session
frequency alone therefore understates the contribution of wider graphs
to request load. Graph width describes structure; execution overlap
depends on request timing.

\label{sourcepara:43}
Session width describes request organization across the session,
whereas intra-task structure describes requests within tasks.
Among tasks in sessions with $W>2$, serial and singleton tasks together
contribute 86.44\% of requests (Table~\ref{tab:wide-session-task-shares}).
Serial tasks still dominate request load in wider sessions.

\label{sourcepara:44}
\begin{table}[t]
  \centering
  \small
  \caption{Request shares by intra-task structure in sessions with $W>2$.}
  \label{tab:wide-session-task-shares}
  \begin{tabular*}{\columnwidth}{@{\extracolsep{\fill}}lccc@{}}
    \toprule
    Intra-task structure & Singleton & Serial & Branching \\
    \midrule
    Request share (\%) & 13.71 & 72.73 & 13.56 \\
    \bottomrule
  \end{tabular*}
\end{table}
\par\nopagebreak[2]
\observationtext{5}{Workflow DAGs are typically small and narrow, yet a minority of sessions with request-graph width greater than one accounts for most requests.}

\label{sourcepara:45}
\noindent\textbf{Workload by intra-task structure.}
Serial tasks carry most request load
(Figure~\ref{fig:structure_summary}(f,g)). They account
for 45.82\% of tasks but contribute 77.62\% of requests and 70.38\%
of cumulative execution duration, measured as the sum of request
execution duration within a task. Execution duration has a long tail within
every structure (Figure~\ref{fig:structure_summary}(f)): P50/P99 are 4.0/138 seconds for singleton,
17.6/588 for serial, and 157/2,525 for branching. Branching tasks are
less frequent but typically execute longer, while serial tasks combine
substantial durations with a much larger population.

\label{sourcepara:46}
Singleton and branching tasks contribute less load overall.
Singleton tasks are common, but each
contains only one request, with shorter inputs and outputs
(Figure~\ref{fig:structure_summary}(j)). Their high frequency
therefore does not translate into a dominant share of request work.

\label{sourcepara:47}
Branching tasks typically contain more requests than serial tasks:
requests per task have P50/P99 of 10/63 for branching and 4/39 for serial
(Figure~\ref{fig:structure_summary}(h)). Requests in both
structures also have higher median cached-input fractions than those
in singleton tasks
(Figure~\ref{fig:structure_summary}(k)) due to a continual request arrival patterns. Input length alone therefore does not indicate how much
prefill work remains~\cite{zheng2024sglang,gao2024cachedattention}.

\label{sourcepara:48}
Branching tasks carry less aggregate load than serial tasks, but their
contribution remains non-negligible: 1.63\% of tasks carry 7.04\%
of requests and 11.21\% of cumulative execution duration
(Figure~\ref{fig:structure_summary}(g)). Their duration
share exceeds their request share, indicating longer mean request
service times than the overall average.
Section~\ref{sec:workflow-execution-dynamics} examines serial continuation
and branching execution.
\par\nopagebreak[2]
\observationtext{6}{Tasks with serial request structures account for 77.6\% of requests despite comprising less than half of all tasks.}
\par

\label{sourcepara:49}
\subsection{Workflow Execution Dynamics}
\label{sec:workflow-execution-dynamics}
\label{sec:inter-task-execution}

\label{sourcepara:50}
Serial chains reveal how request execution and intervening intervals
contribute to workflow duration. Branching graphs raise a complementary
question: whether sibling requests overlap in execution.

\label{sourcepara:51}
\noindent\textbf{Serial continuation.}
Across tasks, continuation gaps run from the predecessor's last request
end to the successor's first request start. Both trigger categories have
median gaps near one second, but P99 reaches 25.72 minutes for
user-triggered successors and 21.65 seconds for agent-triggered successors
(Figure~\label{first:serial_timing}\ref{fig:serial_timing}(a)). These gaps describe when
subsequent work resumes.

\suppressfloats[t]
\begin{figure}[t]
\centering
\includegraphics[width=\columnwidth]{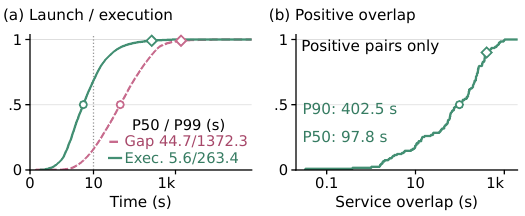}
\setcounter{figure}{10}
\caption{For logical sibling request pairs: (a) Launch gap (dashed) and earlier request duration (solid); (b) Overlap duration among overlapping pairs. Both panels show ECDFs.}
\label{fig:sibling_timing}
\Description{Two distributions show sibling launch gaps and earlier-request durations, and overlap duration for the subset of sibling pairs that overlap.}
\par
\centering
\begin{minipage}{\columnwidth}
  \centering
  \inlinetablecaption{Overlap rates of logical sibling request pairs by launch gap.}
  \label{tab:sibling-overlap-rates}
  \footnotesize
  \setlength{\tabcolsep}{4pt}
  \renewcommand{\arraystretch}{1.02}
  \begin{tabular*}{\columnwidth}{@{\extracolsep{\fill}}lccccc@{}}
    \hline
    Launch gap (s) & $[0,2]$ & $(2,5]$ & $(5,10]$ & $>10$ & Overall \\
    \hline
    Overlap rate (\%) & 1.351 & 0.100 & 0.000 & 0.255 & 0.225 \\
    \hline
  \end{tabular*}
\end{minipage}
\end{figure}

\begin{figure*}[t]
\centering
\includegraphics[width=0.94\textwidth]{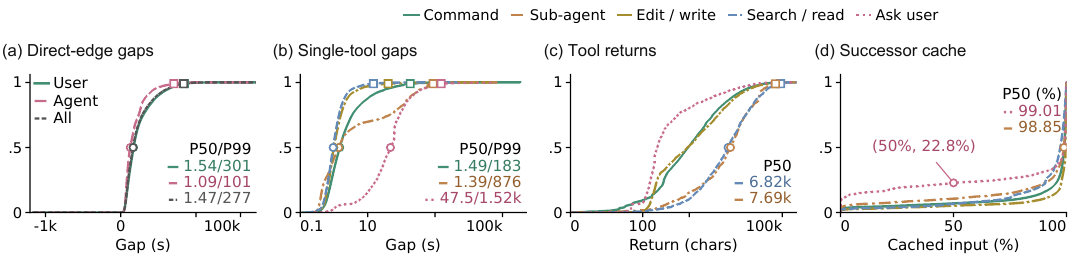}
\setcounter{figure}{11}
\caption{Tool-associated request transitions. ECDFs of (a) direct-edge end-to-start gaps by trigger, (b) single-tool transition gaps, (c) total tool-return characters in intervals exceeding half their serial task's span, grouped by category presence, and (d) first-successor cached-input fractions after single-tool transitions. Panels (b--d) use the five leading categories in Figure~\ref{fig:dominant_interval_tools}.}
\label{fig:execution_tools}
\Description{Four empirical cumulative distributions show direct-edge gaps, single-tool gaps, tool-return characters, and successor cached-input fractions.}
\end{figure*}

\begin{figure}[t]
\setcounter{figure}{12}
\centering
\includegraphics[width=0.88\columnwidth]{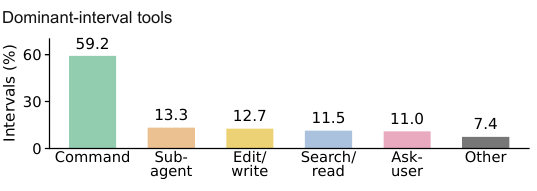}
\caption{Tool presence in intervals exceeding half their serial task's span. Categories can overlap; Other pools planning, MCP, and knowledge search.}
\label{fig:dominant_interval_tools}
\Description{A bar chart shows the percentage of long serial intervals containing command, sub-agent, edit/write, search/read, ask-user, and other tool categories.}
\end{figure}

\label{sourcepara:52}
Within tasks, inter-request intervals contribute to task span and occupy
a larger share in longer serial chains
(Figure~\ref{fig:serial_timing}(b,c)). Here, span
runs from the first request start to the last request end; each interval
runs from one request's end to the next request's start. We compare
per-task mean times for middle cycles. From three-request chains to chains with at least 16
requests, the medians of these task means increase from 1.96 to 5.11
seconds for request intervals and from 4.26 to 6.51 seconds for request execution.
Request execution retains a median span share above 50\%
in every group.

\label{sourcepara:53}
Inter-request time can also concentrate in a single long interval.
A single interval exceeds half the span in 12.01\% of serial tasks,
which account for 41.18\% of cumulative total serial task spans
(Figure~\ref{fig:serial_timing}(d,e)). These tasks therefore
have longer average spans, with more time spent in one interval than
in all request executions combined. This pattern also occurs in 8.8\%
of tasks with 8--15 requests and 5.7\% with at least sixteen, so it
extends beyond short chains. Request execution alone gives
an incomplete view of these tasks' duration.
Section~\ref{sec:request-transitions} examines the tool activities
accompanying these long intervals
(Figure~\label{first:dominant_interval_tools}\ref{fig:dominant_interval_tools}).

\label{sourcepara:54}
\par\nopagebreak[2]
\observationtext{7}{In 12.0\% of serial tasks, one inter-request gap exceeds half the task span; these tasks have longer spans on average.}
\label{observation:serial-long-gaps}
\par

\label{sourcepara:55}
\noindent\textbf{Branching execution.}
Branching tasks carry a non-negligible share of workload and have longer
request executions on average than serial tasks
(Figures~\ref{fig:structure_summary}(g)
and~\ref{fig:structure_summary}(i)). We examine overlap among
logical sibling requests.

\label{sourcepara:56}
Logical sibling requests share a predecessor and have distinct semantic
contexts, but only 0.225\% of pairs overlap
(Table~\ref{tab:sibling-overlap-rates}). Most pairs (83.90\%) launch more
than ten seconds apart (Figure~\label{first:sibling_timing}\ref{fig:sibling_timing}(a)).
Launch spacing must be interpreted relative to request duration: a pair
overlaps only if the later request starts before the earlier one ends.
Even among pairs launched within 2 seconds, only 1.351\% overlap
(Table~\ref{tab:sibling-overlap-rates}). Low overlap is therefore not
confined to widely separated launches; short launch gaps can still
exceed the earlier request's duration.

\label{sourcepara:57}
The few positive overlaps can last for minutes
(Figure~\ref{fig:sibling_timing}(b)). The low rate describes overlap frequency, not duration.
\par\nopagebreak[2]
\observationtext{8}{Only 0.225\% of logical sibling request pairs have overlapping service intervals.}
\label{observation:request-sibling-overlap}
\par

\label{sourcepara:58}
\subsection{Request Transition and Tool Calls}
\label{sec:request-transitions}

\label{sourcepara:59}
Request transitions connect model execution through elapsed time and
continuing context. We summarize their overall characteristics before
examining tool-associated waiting and successor cache reuse.

\label{sourcepara:60}
\noindent\textbf{Requests Transitions Characterizing.}
Inter-request gaps describe when workflow execution continues from one
request to the next. For a direct request edge $u\!\rightarrow\!v$ within
a task, the gap is $g_{uv}=s_v-e_u$, where $s_v$ is the successor's start
time and $e_u$ is the predecessor's end time. Each edge contributes one
observation to the gap distribution. Typical gaps are around one second,
while user-triggered tasks have a longer upper tail
(Figure~\label{first:execution_tools}\ref{fig:execution_tools}(a)). These gaps capture the
intervals between request executions, which can include tool calls and
other waiting, rather than tool runtime alone.

\suppressfloats[t]
\begin{figure}[t]
\centering
\includegraphics[width=0.95\columnwidth]{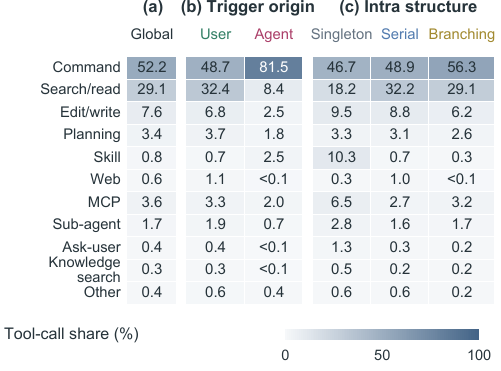}
\setcounter{figure}{13}
\caption{Tool-call composition: (a) global; (b) by trigger origin; (c) by intra-task structure. Columns are independently normalized; global calls and task-attributed calls form separate populations.}
\label{fig:tool_mix}
\Description{A table-like heatmap reports the shares of eleven tool categories globally and within user-triggered, agent-triggered, singleton, serial, and branching groups.}
\end{figure}

\label{sourcepara:61}
\noindent\textbf{Tool-associated waiting.}
We first examine which tool activities accompany the long serial
intervals identified in Section~\ref{sec:workflow-execution-dynamics}.
Command and search/read account for 81.25\% of tool calls overall
(Figure~\label{first:tool_mix}\ref{fig:tool_mix}(a)), and command appears in
59.15\% of intervals exceeding half their task's span (Figure~\ref{fig:dominant_interval_tools}). These
shares describe tool-call frequency and tool presence during long waits
between requests, not time spent executing each tool.

\label{sourcepara:62}
To examine waiting times by tool type, we next consider single-tool
transitions within tasks (Figure~\ref{fig:execution_tools}(b)).
Ask-user has P50/P99 gaps of 47.48/1,515.02 seconds, while sub-agent has
1.39/876.08 seconds. Search/read and edit/write have shorter tails.
Sub-agent continuations usually resume quickly, but some have long waits;
ask-user continuations have longer typical waits.

\label{sourcepara:63}
\noindent\textbf{Information and reuse on continuation.}
For the long serial intervals identified above, tool-return totals
describe the feedback available for continuation. Intervals containing
search/read or sub-agent have higher median return totals than those
containing command or ask-user
(Figure~\ref{fig:execution_tools}(c)). Each total sums all
tool-return characters in the interval.

\label{sourcepara:64}
For single-tool transitions, the first successor request's cached-input
fraction describes how much input is served from existing cached state.
Ask-user and sub-agent successors both have median cached-input fractions
near 99\% (Figure~\ref{fig:execution_tools}(d)), but 22.82\% and
10.68\%, respectively, have less than half their input served from cache.
A larger fraction of ask-user successors therefore process most of their
input without cache reuse, despite the similar medians. Actual reuse
reflects both context continuity and the cached state available when
execution resumes~\cite{gao2024cachedattention,srivatsa2025preble}.

\label{sourcepara:65}
These waiting and reuse differences motivate examining state retention.
Retaining reusable context can avoid repeated input processing on return,
but occupies memory throughout the wait~\cite{abhyankar2024infercept,li2026continuum}. Section~\ref{sec:runtime-performance}
examines this tradeoff through context continuity and state management.

\label{sourcepara:66}
\par\nopagebreak[2]
\observationtext{9}{Waiting-time and successor cache-reuse distributions differ across tool categories.}
\label{observation:tool-transitions}
\par

\label{sourcepara:67}
\section{Infrastructure-level Serving Demands}
\label{sec:runtime-performance}

\label{sourcepara:68}
Request volume alone does not capture the serving demands of these
workflows. We examine how state and token work concentrate across
sessions, how context continuity defines cache affinity, and why
heterogeneous waits motivate dynamic state retention.

\label{sourcepara:69}
\subsection{Concentrated Serving Demand}
\label{sec:infra-concentration}

\label{sourcepara:70}
\noindent\textbf{State and token demand.}
Token work and context residency concentrate in a small fraction of
sessions. We measure input processing by \emph{new-prefill tokens}, the
input tokens not served from cache, and output generation by
\emph{decode tokens}, the generated output tokens.
\emph{Context residency} measures logical context size accumulated over time:
\begingroup
\setlength{\abovedisplayskip}{0pt}
\setlength{\belowdisplayskip}{0pt}
\setlength{\abovedisplayshortskip}{0pt}
\setlength{\belowdisplayshortskip}{0pt}
\[
 L_s=\int_{t_s^{\mathrm{start}}}^{t_s^{\mathrm{end}}} K_s(t)\,dt,
\]
\endgroup
where $K_s(t)$ is the session's logical context size in tokens.
The integral runs from session start to the last request's end and
gives a logical proxy for cumulative state demand in token-seconds.
KV-cache memory occupancy depends on cached state
and its model-specific storage cost~\cite{kwon2023pagedattention}.

\label{sourcepara:71}
Each demand is highly concentrated
(Figure~\label{first:infra_demand}\ref{fig:infra_demand}(a)). To identify sessions with high
demand across all three measures, we intersect the top 20\% of sessions
ranked independently by each measure.
These \emph{selected sessions} comprise 14.3\% of all sessions but account
for 98.5\% of context residency, 92.3\% of new-prefill tokens, and 86.0\%
of decode tokens. The three demands therefore concentrate in a common
group, which we examine in the subsequent cache-affinity analysis and
retention model (\S\ref{sec:infra-retention-model}).

\suppressfloats[t]
\begin{figure}[t]
\centering
\includegraphics[width=\columnwidth]{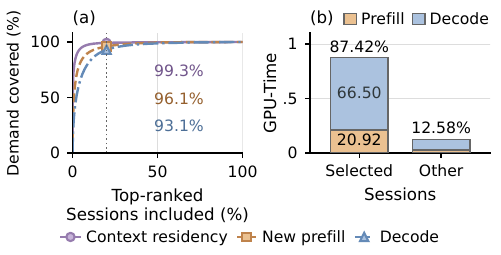}
\setcounter{figure}{14}
\caption{State and token demand. (a) Independent session rankings by demand, with top-20\% coverage marked. (b) Shares of total cumulative request engine duration by session group and prefill/decode phase.}
\label{fig:infra_demand}
\Description{Two panels compare concentration of inference demand across sessions and the execution-time breakdown among high-demand sessions.}
\end{figure}

\begin{figure*}[t]
\centering
\includegraphics[width=\textwidth]{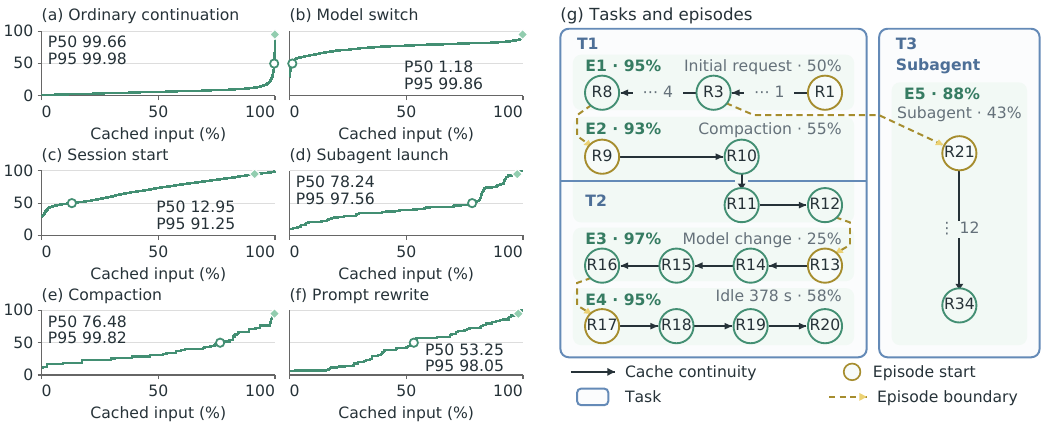}
\setcounter{figure}{15}
\caption{Context continuity motivates affinity boundaries that need not align with tasks. (a--f) Input-token-weighted ECDFs of per-request cached-input percentage by context event; circles/diamonds mark P50/P95. (g) Task outlines and episode backgrounds share request nodes; green/gray percentages give continuation/start reuse. Task outlines and fragment links are schematic; the parent--subagent edge is observed. E4 uses a 300-second idle boundary; ellipses count omitted requests.}
\label{fig:cache_episodes}
\Description{Six empirical cumulative distributions relate cache reuse to context events and idle intervals; a workflow schematic illustrates episode boundaries within a session.}
\end{figure*}

\label{sourcepara:72}
\noindent\textbf{Service-time composition.}
The selected sessions also dominate cumulative request engine duration,
and the decode phase accounts for 76.1\% of their duration
(Figure~\ref{fig:infra_demand}(b)). We sum request start-to-end
intervals, counting overlaps separately, and use TTFT and the remaining
request duration for the prefill and decode phases, respectively.
Both phase durations include waiting and scheduling overheads.

\label{sourcepara:73}
\par\nopagebreak[2]
\observationtext{10}{A common small group of sessions accounts for most token work and logical context residency, with decode accounting for most cumulative request service time within this group.}

\label{sourcepara:74}
\subsection{Cache Affinity Episode Within Sessions}
\label{sec:infra-episodes}

\label{sourcepara:75}
After characterizing demand across sessions, we examine context
continuity among requests within sessions and its relation to task
boundaries.

\label{sourcepara:76}
\noindent\textbf{Boundaries for state management.}
Tasks organize semantic progress, while serving affinity concerns the
historical context that successive requests can reuse. Characterizing
this affinity requires considering both changes to that context and the
time over which it is retained~\cite{zheng2024sglang,li2026continuum}. We examine context events and
idle intervals as complementary signals for identifying state-management
boundaries within sessions.

\label{sourcepara:77}
\noindent\textbf{Episodes and observed cache affinity.}
To make these context and retention boundaries explicit, we introduce
an \emph{episode} as a serving-affinity unit: a group of requests that
continue one context history within a retention horizon. It captures
continuity of potentially reusable state across requests and their
intervening waits, allowing its boundaries to differ from task
boundaries. An episode starts with the first observed request in a
context history or with a context-changing event: a model switch,
subagent launch, compaction, or system-prompt rewrite. Requests that
continue this history remain in the episode while the idle interval
from the preceding request's end to the next request's start does not
exceed the retention horizon; after a longer interval, the next request
starts a new episode. Parent and subagent histories are tracked
separately, so a subagent launch starts the subagent's episode without
itself ending the parent's. Context events mark changes in the history
being continued; the chosen retention horizon determines which idle
intervals split episodes. Figure~\label{first:cache_episodes}\ref{fig:cache_episodes}(g) illustrates
this grouping with a 300-second horizon. This threshold is a grouping
parameter, not a measured eviction time or an optimal TTL.

\label{sourcepara:78}
Within sessions, we measure token reuse as $R=\sum_r C_r/\sum_r I_r$,
where $I_r$ and $C_r$ count input and cached-input tokens. Ordinary
continuations reach 93\% measured reuse with the same model and serving
instance and short idle intervals. The input-token-weighted distributions
concentrate near full reuse for ordinary continuations, while model
switches and session starts place substantial mass near zero
(Figure~\ref{fig:cache_episodes}(a--c)). Subagent launches, compaction,
and prompt rewrites have broader distributions but retain high-reuse
upper tails, with P95 above 97\% (Figure~\ref{fig:cache_episodes}(d--f)). These distributions
show that reusable state can persist across potential affinity
boundaries. As with post-tool reuse (Figure~\ref{fig:execution_tools}(d)),
event type alone does not determine actual cache reuse. Reuse also weakens
with longer idle intervals, measured from the preceding request's end to
the next request's start, although high-reuse continuations remain after
long waits (Figure~\label{first:idle_reuse}\ref{fig:idle_reuse}).

\begin{figure}[t]
\centering
\includegraphics[width=0.92\columnwidth]{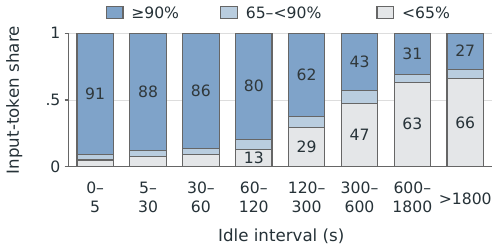}
\setcounter{figure}{16}
\caption{Cache reuse by idle interval for same-model, same-instance continuations. Bars show input-token shares by reuse band; labels mark low- and high-reuse percentages.}
\label{fig:idle_reuse}
\Description{Cache reuse is compared across idle intervals between consecutive requests.}
\end{figure}

\label{sourcepara:79}
Task and episode boundaries differ. Cross-task continuations within an
episode retain 80.59\% actual token reuse, compared with 44.67\% for
same-task continuations across episodes
(Table~\ref{tab:infra-task-episode}). Figure~\ref{fig:cache_episodes}(g)
illustrates both cases: E2 spans T1/T2, while context changes split
episodes within a task. Task progress and cache continuity thus follow
different boundaries.

\label{sourcepara:80}
\begin{table}[t]
\centering
\begin{minipage}{\columnwidth}
  \centering
  \inlinetablecaption{Adjacent-request pair shares and token reuse by task and episode boundaries.}
  \label{tab:infra-task-episode}
  \small
  \setlength{\tabcolsep}{3pt}
  \begin{tabular*}{\columnwidth}{@{\extracolsep{\fill}}llrrr@{}}
    \hline
    \textbf{Task} & \textbf{Episode} & \textbf{Pair share} & \multicolumn{2}{c}{\textbf{Token reuse (\%)}} \\
    & & (\%) & \textbf{Actual} & \textbf{Upper bound} \\
    \hline
    Same & Same & 80.91 & 90.08 & 96.81 \\
    Same & Different & 1.66 & 44.67 & 71.98 \\
    Different & Same & 14.86 & 80.59 & 91.85 \\
    Different & Different & 2.58 & 41.20 & 86.95 \\
    \hline
  \end{tabular*}
\end{minipage}
\end{table}

\label{sourcepara:81}
\noindent\textbf{Reusable context and management granularity.}
The \emph{reuse upper bound} replaces cached-input counts in $R$ with
potentially reusable prefix tokens computed from the preceding and
current contexts, assuming that the reusable state remains fully
available. High upper bounds across episode boundaries show that
context changes need not eliminate reusable prefixes. The gap between
these bounds and actual reuse distinguishes content-level reuse
potential from cached-state availability under placement and eviction~\cite{srivatsa2025preble}.
Episodes group requests by context and retention continuity, providing a
concrete basis for examining serving affinity. These observations motivate other resource-management granularities and
policies that account for reusable state beyond task boundaries.

\label{sourcepara:82}
\observationtext{11}{Context reuse can span tasks while context can change within a task, motivating episodes as serving-affinity units that group requests by context and retention continuity.}

\label{sourcepara:83}
\subsection{Dynamic State Retention}
\label{sec:infra-retention}

\label{sourcepara:84}
Context continuity identifies potentially reusable state; a time-to-live
(TTL) bounds its retention during a wait. Among single-tool transitions,
gaps exceed 300 seconds for 0.7\% of command
transitions and about 7\% each for sub-agent and ask-user
(Figure~\ref{fig:execution_tools}(b)). A common retention
horizon covers different fractions of observed returns
across waiting types.

\label{sourcepara:85}
Retaining reusable context can avoid recomputation on return, but
occupies memory throughout the wait. TTL choices therefore depend on
reusable context, return timing, and available memory. InferCept,
Continuum, and SAGA guide retention using recovery costs, return
estimates, or memory pressure~\cite{abhyankar2024infercept,li2026continuum,guo2026saga}.
CacheWise likewise uses workload characteristics to guide KV-cache
management for coding agents~\cite{tiwari2026cachewise}.
Section~\ref{sec:infra-retention-model} quantifies this tradeoff under
explicit retention budgets and discusses the information needed online.

\label{sourcepara:86}
\observationtext{12}{TTL decisions should adapt to waiting type, reusable context, and available resources: extending retention trades memory occupied during a wait against recomputation on return.}

\label{sourcepara:87}
\section{Implications and Open Questions}
\label{sec:open-questions}

\label{sourcepara:88}
We discuss cross-layer abstraction, episode-based scheduling, and adaptive
KV cache management, relating each direction to its benefits, costs, and
information needs at runtime. An offline model quantifies the opportunity
for allocating retention across waiting types.

\label{sourcepara:89}
\subsection{Cross-Layer Abstraction}
\label{sec:implications-task-semantics}
\label{sec:implications-workflow-execution}

\label{sourcepara:90}
Task-level initiation semantics and workflow-level execution patterns provide information that
could guide capacity allocation and request scheduling
(Sections~\ref{sec:task-initiation}--\ref{sec:workflow-structure-execution}).
Structured abstractions already allow applications to expose some of this
information: Parrot uses Semantic Variables to represent relationships
among requests and make them available to the serving
system~\cite{lin2024parrot}. These mechanisms raise a further question:
which dynamic signals provide useful information beyond the requests
already queued or running, and when must they become available to affect
a decision?

\label{sourcepara:91}
The value of a signal depends on the decision it supports. Expected task
activations could inform capacity planning before requests arrive, while
execution readiness and progress could inform scheduling as a workflow
unfolds~\cite{luo2026agentix}. The observed differences across triggering categories and
workflow patterns motivate examining these signals, but their additional
value depends on what the infrastructure can already infer from request
histories and current load. A useful signal must arrive before the
relevant decision and remain informative as execution changes.

\label{sourcepara:92}
A cross-layer abstraction could represent these signals as hints that
applications, orchestrators, and serving systems interpret consistently~\cite{zheng2024sglang}.
Execution constraints need to be distinguished from estimates of future
work, and updates need to reflect changes in the workflow~\cite{kim2024llmcompiler,zheng2026pbkv}. The research
question is which hints improve resource and scheduling decisions under
these conditions, including when estimates are inaccurate or updates
arrive late.

\label{sourcepara:93}
\openquestiontext[0]{1}{How can dynamic information from task initiation and workflow execution be selected and exposed to improve inference resource management and scheduling beyond request-level observations?}

\label{sourcepara:94}
\subsection{Episode-Based Scheduling}
\label{sec:infra-episode-placement}

\label{sourcepara:95}
Context continuity can span tasks or change within a task
(Section~\ref{sec:infra-episodes}), making task boundaries alone
insufficient to describe serving affinity. Episodes make these changes
explicit and provide opportunities to reconsider placement. For
episode-based scheduling, the question is how context continuity should
guide the grouping of requests for placement and the timing of reassignment.

\label{sourcepara:96}
A context boundary does not by itself determine whether placement should
change. Shared prefixes may remain cached across a boundary, while
resource pressure may justify reassignment during a continuing episode.
The appropriate affinity unit therefore depends on both the state that
can be reused and the work that remains. A policy needs to weigh the
expected reduction in queueing and execution time against the cost of
transferring state or rebuilding context elsewhere~\cite{srivatsa2025preble,sun2024llumnix,liu2024cachegen}.

\label{sourcepara:97}
These quantities evolve during execution. Current load and cache
availability describe immediate conditions, while request progression
and tool activity may help estimate future work. Overestimating that
work can trigger a move whose cost is never recovered; underestimating
it can prolong imbalance. The challenge is to determine when available
evidence justifies changing affinity and when preserving the current
assignment is preferable. Retention affects this decision through the
state that remains available, but does not prescribe where subsequent
requests should execute~\cite{qin2025mooncake}.

\label{sourcepara:98}
\openquestiontext[0]{2}{How should episode-based scheduling adapt placement granularity and reassignment timing to context continuity and uncertain remaining work?}

\label{sourcepara:99}
\subsection{Adaptive KV Cache Management}
\label{sec:infra-retention-model}

\label{sourcepara:100}
Waiting-time differences and persistent context reuse
(Sections~\ref{sec:infra-episodes}--\ref{sec:infra-retention}) motivate
adapting retention to expected reuse and memory pressure.

\begin{figure}[t]
\centering
\includegraphics[width=\columnwidth]{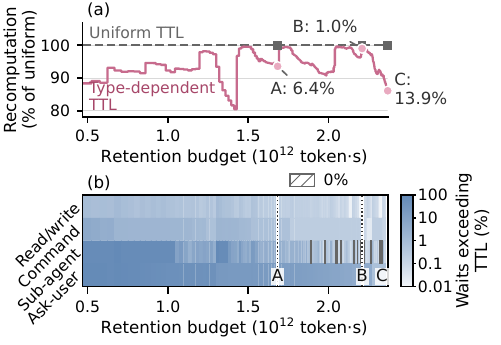}
\setcounter{figure}{17}
\caption{Offline retention allocation. (a) Remaining recomputation normalized to uniform TTL at each budget (100\%); annotations show relative savings. (b) Within-type shares of waits exceeding allocated TTLs; hatching denotes zero. A--C mark budgets from uniform 5-, 15-, and 25-minute TTLs.}
\label{fig:retention}
\Description{Two panels compare cache retention and recomputation tradeoffs across waiting types and retention budgets.}
\end{figure}

\label{sourcepara:101}
\noindent\textbf{An offline allocation model.}
Using the selected 14.3\% of sessions, we model returning waits across $K=4$ types:
read/write (search/read and edit/write), command, sub-agent, and ask-user.
For type $c$ with TTL $\tau_c$, retention cost $C_c(\tau_c)$ sums each
source request's input-context size times its retained duration, from
completion until the next request starts or the TTL expires.
Recomputation cost $U_c(\tau_c)$ sums source-input tokens for returns
after expiry. These measure cumulative logical retention in token-seconds
and modeled recomputation in tokens, respectively.
Following the shared-resource view of cache allocation~\cite{dehghan2016utility},
we minimize recomputation under a retention budget $B$:
\begingroup
\setlength{\abovedisplayskip}{0pt}
\setlength{\belowdisplayskip}{0pt}
\setlength{\abovedisplayshortskip}{0pt}
\setlength{\belowdisplayshortskip}{0pt}
\[
\begin{array}{l}
 V(B)=\displaystyle\min_{\tau_1,\ldots,\tau_K\ge0}\sum_{c=1}^K U_c(\tau_c),\\[3pt]
 \mathrm{subject\ to}\quad \displaystyle\sum_{c=1}^K C_c(\tau_c)\le B.
\end{array}
\]
\endgroup
The model selects one fixed TTL per type; the baseline uses the best
feasible uniform TTL, with recomputation $U_{\mathrm{uniform}}(B)$.

\label{sourcepara:102}
\noindent\textbf{Savings at matched retention cost.}
Uniform 5-, 15-, and 25-minute TTLs define three reference budgets;
type-dependent TTLs are allocated separately. The reduction in remaining
recomputation,
$(U_{\mathrm{uniform}}-V)/U_{\mathrm{uniform}}$, is 6.4\%, 1.0\%, and
13.9\%, respectively (Figure~\label{first:retention}\ref{fig:retention}(a)).
The middle setting is close to uniform, so gains are not uniformly
large or monotonic in budget.

\label{sourcepara:103}
Longer waiting tails need not receive longer TTLs: ask-user receives
a shorter TTL than sub-agent at the 5-minute reference budget, but a
longer TTL at the 25-minute budget. Allocation depends on
recomputation avoided per added retention cost, accounting for context
size and return timing, rather than equalizing the fractions of waits
beyond TTL (Figure~\ref{fig:retention}(b)).

\label{sourcepara:104}
\noindent\textbf{Online memory allocation.}
Online retention faces an instantaneous memory limit, whereas the model
constrains cumulative retention cost. Waiting contexts compete with
active requests for memory, so retaining one context can avoid
recomputation but delay other
work~\cite{abhyankar2024infercept,kwon2023pagedattention}.
Retention decisions need estimates of whether a context will return,
when it will return, and how much reusable state it
carries~\cite{zheng2026pbkv,li2026continuum,yu2025pensieve}.
Waiting type, execution progress, context changes, and completion signals
can update these estimates~\cite{gao2024cachedattention}. The challenge
is to revise retention as reuse estimates and resource pressure change,
including for contexts that never return.

\label{sourcepara:105}
\openquestiontext[0]{3}{How should adaptive KV cache management allocate memory between waiting contexts and active execution as reuse estimates and resource demand change?}

\label{sourcepara:106}
\section{Other Related Work}
\label{sec:related-work}
Beyond workload characterization, prior work studies efficient agent execution
and inference serving.
Agent methods and frameworks support reasoning, tool use, and
collaboration~\cite{wu2024autogen,zhuge2024gptswarm,zhang2025aflow,kim2024llmcompiler,yao2023react,schick2023toolformer,shinn2023reflexion,hong2024metagpt}.
Benchmarks and experimentation frameworks evaluate execution costs
and serving performance~\cite{chang2026agentsysbench,kondrashov2026aries}.
LLM serving systems optimize computation, batching, and resource
use~\cite{kwon2023pagedattention,dao2022flashattention,leviathan2023speculative,yu2022orca,agrawal2024sarathi,zhong2024distserve,patel2024splitwise,yu2026faascale,zhang2026janusdisaggregatingattentionexperts}.
Execution and scheduling systems exploit program structure, request
dependencies, and context
locality~\cite{zheng2024sglang,lin2024parrot,luo2026agentix,guo2026saga,srivatsa2025preble,sun2024llumnix,wang2026smetric}.
KV-cache systems support context reuse through retention, transfer,
and storage~\cite{tiwari2026cachewise,gao2024cachedattention,yu2025pensieve,abhyankar2024infercept,li2026continuum,zheng2026pbkv,liu2024cachegen,qin2025mooncake}.
Our end-to-end trace analysis complements these efforts with production
evidence to inform serving optimizations such as request scheduling and state management.

\label{sourcepara:107}
\section{Conclusion}

\label{sourcepara:108}
We characterize agentic workloads on a large-scale production platform
with over 10k GPUs, using a two-week trace of 11.7 million sampled requests
across 948.4k sessions.
Through an analysis of task-level initiation semantics, workflow-level
execution patterns, and infrastructure-level serving demands, we examine
how task and workflow behavior relates to request load, execution timing,
and context reuse.
Our observations motivate further research on cross-layer abstractions,
episode-based scheduling, and adaptive KV-cache management for agent serving.
We plan to release sanitized traces from our characterization data to the public soon.
\par

\setlength{\bibsep}{0pt}
\enlargethispage{3pt}
\bibliographystyle{unsrt}
\bibliography{references}

\end{document}